\input amstex
\documentstyle{amsppt}
%
%
\catcode`@=11
\redefine\output@{%
  \def\break{\penalty-\@M}\let\par\endgraf
  \ifodd\pageno\global\hoffset=105pt\else\global\hoffset=8pt\fi  
  \shipout\vbox{%
    \ifplain@
      \let\makeheadline\relax \let\makefootline\relax
    \else
      \iffirstpage@ \global\firstpage@false
        \let\rightheadline\frheadline
        \let\leftheadline\flheadline
      \else
        \ifrunheads@ 
        \else \let\makeheadline\relax
        \fi
      \fi
    \fi
    \makeheadline \pagebody \makefootline}%
  \advancepageno \ifnum\outputpenalty>-\@MM\else\dosupereject\fi
}
\catcode`\@=\active
\nopagenumbers
\TagsOnRight
\def\Orth{\operatorname{O}}
\def\so{\operatorname{so}}
\def\SO{\operatorname{SO}}
\def\EM#1#2{O\raise 1pt
\hbox{$\ssize\thickfracwithdelims[]\thickness0{#1}{#2}$}}
\def\SM#1#2{S\raise 1pt
\hbox{$\ssize\thickfracwithdelims[]\thickness0{#1}{#2}$}}
\def\negskp{\hskip -2pt}
\def\blue#1{#1}

\catcode`#=11\def\diez{#}\catcode`#=6
\catcode`_=11\def\podcherkivanie{_}\catcode`_=8
\def\mycite#1{\cite{\blue{#1}}\immediate\special{ps:
     ShrHPSdict begin /ShrBORDERthickness 0 def}}
\def\myciterange#1#2#3#4{\cite{\blue{#2#3#4}}\immediate\special{ps:
     ShrHPSdict begin /ShrBORDERthickness 0 def}}
\def\mytag#1{%
    \tag#1}
\def\mythetag#1{\thetag{\blue{#1}}\immediate\special{ps:
     ShrHPSdict begin /ShrBORDERthickness 0 def}}
\def\myrefno#1{\no#1}
\def\myhref#1#2{\blue{#2}\immediate\special{ps:
     ShrHPSdict begin /ShrBORDERthickness 0 def}}

\def\mytheorem#1{\csname proclaim\endcsname{Theorem #1}}
\def\mytheoremwithtitle#1#2{\csname proclaim\endcsname{Theorem #1#2}}
\def\mythetheorem#1{\blue{#1}\immediate\special{ps:
     ShrHPSdict begin /ShrBORDERthickness 0 def}}
\def\mylemma#1{\csname proclaim\endcsname{Lemma #1}}
\def\mylemmawithtitle#1#2{\csname proclaim\endcsname{Lemma #1#2}}

\def\mycorollary#1{\csname proclaim\endcsname{Corollary #1}}

\pagewidth{360pt}
\pageheight{606pt}
\topmatter
\rightheadtext{Algorithm for generating orthogonal matrices.}
\topmatter
\title Algorithm for generating orthogonal matrices
with rational elements.
\endtitle
\author
R.~A.~Sharipov
\endauthor
\address 5 Rabochaya street, 450003 Ufa, Russia\newline
\vphantom{a}\kern 12pt Cell Phone: +7(917)476 93 48
\endaddress
\email \vtop to 30pt{\hsize=280pt\noindent
\myhref{mailto:r-sharipov\@mail.ru}
{r-sharipov\@mail.ru}\newline
\myhref{mailto:R\podcherkivanie Sharipov\@ic.bashedu.ru}
{R\_\hskip 1pt Sharipov\@ic.bashedu.ru}\vss}
\endemail
\urladdr
\vtop to 20pt{\hsize=280pt\noindent
\myhref{http://www.freetextbooks.narod.ru}
{http:/\negskp/www.freetextbooks.narod.ru}\newline
\myhref{http://sovlit2.narod.ru/}
{http:/\negskp/sovlit2.narod.ru}\vss}
\endurladdr
\abstract
Special orthogonal $n\times n$ matrices with rational elements form the 
group $\SO(n,\Bbb Q)$, where $\Bbb Q$ is the field of rational numbers.
A theorem describing the structure of an arbitrary matrix from this group
is proved. This theorem yields an algorithm for generating such matrices
by means of random number routines.
\endabstract
\endtopmatter
\loadbold
\document

\head
1. Introduction.
\endhead
    Orthogonal matrices from the orthogonal group $\Orth(n,\Bbb R)$ describe
rotations and rotations with reflections in the $n$-dimensional
Euclidean space $\Bbb R^n$. Pure rotations constitute another
classical group $\SO(n,\Bbb R)$, which is a subgroup in $\Orth(n,\Bbb R)$.
The following matrix represents an elementary rotation in the $p$-$q$ 
coordinate plane:
\vskip 2pt plus 2pt
\leftline{\hskip 129pt\vrule height 0pt depth 3pt
\vrule height 0pt depth 1pt width 13pt
\lower 3pt\hbox{\ $p$}
\vrule height 0pt depth 1pt width 13pt
\vrule height 0pt depth 3pt}
\vskip -4pt
$$
\hskip -2em
\EM{p}{q}(\varphi)=\Vmatrix 1 &\hdots &0 &\hdots &0 &\hdots &0\\
\vdots &\ddots &\vdots &\ &\vdots &\ &\vdots\\
0 &\hdots &\alpha &\hdots &-\beta &\hdots &0\\
\vdots &\ &\vdots &\ddots &\vdots &\ &\vdots\\
0 &\hdots &\beta &\hdots &\alpha &\hdots &0\\
\vdots &\ &\vdots &\ &\vdots &\ddots &\vdots\\
0 &\hdots &0 &\hdots &0 &\hdots &1
\endVmatrix.
\mytag{1.1}
$$
\vskip 1pt
\leftline{\hskip 129pt\vrule height 3pt depth 1pt
\vrule height 0pt depth 1pt width 34pt
\lower 3pt\hbox{\ $q$}
\vrule height 0pt depth 1pt width 34pt
\vrule height 3pt depth 1pt}
\vskip 8pt plus 2pt
\noindent Here $\alpha=\cos(\varphi)$, $\beta=\sin(\varphi)$, and
$\varphi$ is the rotation angle. Matrices of the form \mythetag{1.1}
can be treated as elements of $\SO(2,\Bbb R)$ embedded into
$\SO(n,\Bbb R)$. The matrix
$$
\hskip -2em
\Omega=\Vmatrix
\lower 12pt\hbox{\vbox{\hrule\hbox to 40pt{\vrule height
18pt depth 13pt\hss $O^*$\hss
\vrule height 18pt depth 13pt}\hrule}} &
\matrix 0\\ \vdots\\ 0\endmatrix\\
\vspace{5pt}
\matrix 0 &\hdots &0\endmatrix & 1
\endVmatrix
\mytag{1.2}
$$
with $O^*\in\SO(n-1,\Bbb R)$ then is an element of $\SO(n-1,
\Bbb R)$ embedded into $\SO(n,\Bbb R)$. For orthogonal matrices
over the field of reals $\Bbb R$ there is the following theorem.
\mytheorem{1.1} Each orthogonal matrix $O\in\SO(n,\Bbb R)$
can be represented as a product $O=\EM{1}{2}(\varphi_1)\cdot\ldots
\cdot\EM{n-1}{n}(\varphi_{n-1})\cdot\Omega$, where the first $n-1$
terms are matrices of elementary rotations \mythetag{1.1}, and
$\Omega$ is an orthogonal matrix of the form \mythetag{1.2}. 
\endproclaim
    The angles $\varphi_1,\,\ldots,\,\varphi_{n-1}$ in the 
theorem~\mythetheorem{1.1} are known as Euler angles (see
Chapter~\uppercase\expandafter{\romannumeral 7} in \mycite{1} for
Euler angles in the three-dimensional case $n=3$). These angles are 
restricted by the inequalities
$$
\xalignat 2
&0\leqslant\varphi_1\leqslant 2\pi,
&&0\leqslant\varphi_i\leqslant\pi\text{\ \ for \ }i=2,
\,\ldots,\,n-1.\quad
\endxalignat
$$
Applying the theorem~\mythetheorem{1.1} recursively to $O$, then to $O^*$ 
in \mythetag{1.2}, and so on, we easily prove the following theorem.
\mytheorem{1.2} Each orthogonal matrix $O\in\SO(n,\Bbb R)$
can be represented as a product of $n(n-1)/2$ matrices of elementary
rotations \mythetag{1.1}.
\endproclaim
\noindent The theorem~\mythetheorem{1.1} is not valid for orthogonal matrices 
over the field of rational numbers. As for the theorem~\mythetheorem{1.2}, 
I don't know if it is valid or not for $O\in\SO(n,\Bbb Q)$. However, there is 
an algorithm for constructing orthogonal matrices over the field of rational 
numbers. This algorithm is exhausting, i\.\,e\. each orthogonal matrix 
$O\in\SO(n,\Bbb Q)$ can be obtained by applying this algorithm.\par
\head
2. Stereographic projection.
\endhead
\parshape 19 0pt 360pt 0pt 360pt 0pt 360pt
180pt 180pt 180pt 180pt 180pt 180pt 180pt 180pt 180pt 180pt
180pt 180pt 180pt 180pt 180pt 180pt 180pt 180pt 180pt 180pt
180pt 180pt 180pt 180pt 180pt 180pt 180pt 180pt 180pt 180pt
0pt 360pt
    Let $\goth S$ be the unit sphere in $\Bbb R^n$. We take the point $S$
with the coordinates $(0,\ldots,0,-1)$ for the south pole on this sphere.
\vadjust{\vskip 0pt\hbox to 0pt{\kern -10pt
\includegraphics{f13-2.eps}\hss}\vskip 0pt}Then the equatorial
hyperplane $\alpha$ is given by the equation $x^n=0$,
where $x^n$ is the $n$-th coordinate of a point of $\Bbb R^n$ (we use
upper indices for coordinates of vectors and points according to
Einstein's tensorial notation, which is popular in differential
geometry and in general relativity). Let $X$ be an arbitrary point
of the unit sphere $\goth S$ and let $\left[SX\right>$ be the ray
starting at the south pole $S$ and passing through the point
$X\in\goth S$. This ray crosses the equatorial hyperplane at some 
definite point $Y$ (as shown on figure~2.1 for three dimensional case). 
For each $X$ point $Y$ is unique. If we denote by $\goth S^\circ$ the 
unit sphere $\goth S$ with the pinned off south pole $S$, then we get 
the mapping $f\!:\goth S^\circ\to\alpha$ that maps $\goth S^\circ$ onto 
equatorial hyperplane $\alpha$. This map is called the {\it stereographic 
projection}. It is bijective and smooth. It's very important for us that 
the stereographic projection is given by rational functions. Indeed, if 
$x^1,\,\ldots,\,x^n$ are the coordinates of some point $X\in\goth S^\circ$ 
and if $y^1,\,\ldots,\,y^{n-1}$ are the coordinates of the point $Y=f(X)$, 
then for $y^1,\,\ldots,\,y^{n-1}$, we get the following formulas:
$$
\hskip -2em
y^s=\frac{x^s}{1+x^n}\text{\ \ for \ }s=1,\,\ldots,\,n-1.
\mytag{2.1}
$$
The inverse mapping $f^{-1}\!:Y\to X$ is also given by rational functions.
Indeed, for the $n$-th coordinate $x^n$ we have the formula 
$$
\hskip -2em
x^n=\left.\left(1-\shave{\sum^{n-1}_{i=1}}(y^i)^2\right)\right/
\left(1+\shave{\sum^{n-1}_{i=1}}(y^i)^2\right).
\mytag{2.2}
$$
For the other coordinates of the point $X$ from \mythetag{2.1} and 
\mythetag{2.2} we derive
$$
\hskip -2em
x^s=\frac{2\,y^s}{\left(1+\shave{\sum^{n-1}_{i=1}}(y^i)^2\right)}
\text{, \ where \ }s=1,\,\ldots,\,n-1.
\mytag{2.3}
$$
Rationality of the expressions \mythetag{2.1}, \mythetag{2.2}, and
\mythetag{2.3} mean that rational points of the pinned sphere $\goth S^\circ$
are in one-to-one correspondence with rational points of the equatorial 
hyperplane $\alpha$. As for the south pole $S$, it is usually associated 
with the infinite point $Y=\infty$ on $\alpha$. Indeed, passing to the 
limit $Y\to\infty$ in the formulas \mythetag{2.2} and \mythetag{2.3}, one 
can get the coordinates of the point $S$.
\head
3. Orthogonal matrices and ONB's.
\endhead
    Let $O\in\SO(n,\Bbb Q)$ be some orthogonal matrix. Its columns
can be treated as vectors in $\Bbb Q^n$. Let's denote them $\bold e_1,
\,\ldots,\,\bold e_n$. Orthogonality of $O$ means that the transposed
matrix $O^{\sssize T}$ coincides with the inverse matrix $O^{-1}$:
$$
\hskip -2em
O\cdot O^{\sssize T}=1.
\mytag{3.1}
$$
Written in explicit form the equality \mythetag{3.1} means that 
$\bold e_1,\,\ldots,\,\bold e_n$ are vectors of the unit length
perpendicular to each other. They form a so called ONB (orthonormal
basis) in $\Bbb Q^n$ with respect to the standard scalar product
$$
(\bold X,\,\bold Y)=\sum^n_{i=1}x^i\,y^i.
$$
Matrices from the special orthogonal group $\SO(n,\Bbb Q)$ obey
the additional restriction
$$
\hskip -2em
\det Q=1.
\mytag{3.2}
$$
For the basis vectors $\bold e_1,\,\ldots,\,\bold e_n$ the restriction 
\mythetag{3.2} means that $\bold e_1,\,\ldots,\,\bold e_n$ form a {\it 
right oriented} ONB (or a {\it right} ONB). In three dimensional space 
$\Bbb Q^3$ this property can be visualized.
\definition{Orientation rule} Vectors $\bold e_1,\,\bold e_2,\,\bold
e_3$ form a {\it right triple} in $\Bbb Q^3$ if when looking from the
end point of the third vector $\bold e_3$ the shortest rotation from
the first vector $\bold e_1$ to the second vector $\bold e_2$ is observed 
as a counterclockwise rotation.
\enddefinition
\noindent In physics (in electricity and magnetism) this rule is
formulated as the rule of a right screw and also as the rule of a 
left hand (see \cite{2} and \cite{3}). In multidimensional spaces $n>3$
one cannot visualize the concept of left and right since a human
has no visual experience of living in such spaces. However, one can 
understand it mathematically by means of the theory of
determinants and skew-symmetric polylinear forms. Indeed, one
should first prescribe right orientation to the standard basis in
$\Bbb Q^n$ composed by the vectors
$$
\hskip -2em
\bold E_1=\Vmatrix 1\\ 0\\ \vdots\\ 0\endVmatrix,
\quad\bold E_2=\Vmatrix 0\\ 1\\ \vdots\\ 0\endVmatrix,
\quad\ldots,\quad
\bold E_n=\Vmatrix 0\\ 0\\ \vdots\\ 1\endVmatrix,
\mytag{3.3}
$$
then one should consider the transition matrix $S$ relating the 
standard basis \mythetag{3.3} with another basis $\bold e_1,\,\ldots,\,
\bold e_n$. Its components are defined by the equality
$$
\hskip -2em
\bold e_i=\sum^n_{j=1}S^j_i\cdot\bold E_j,\qquad i=1,\,\ldots,\,n.
\mytag{3.4}
$$
\definition{Definition 3.1} A basis $\bold e_1,\,\ldots,\,\bold e_n$
in $\Bbb Q^n$ is called {\it right oriented} if $\det S>0$. Otherwise,
if $\det S<0$, this base is called {\it left oriented}.
\enddefinition
     Another method of defining orientation in $\Bbb Q^n$ uses skew-symmetric
polylinear forms. Remember that a polylinear $n$-form in $\Bbb Q^n$ is
a $\Bbb Q$-numeric function with $n$ vectorial arguments $\omega=
\omega(\bold X_1,\ldots,\bold X_n)$ which is linear with respect to
each its argument. A form $\omega$ is called completely skew-symmetric
if its value changes the sign under permutation of any pair of arguments
in it. It is known that a completely skew-symmetric $n$-form in $\Bbb Q^n$ 
is determined uniquely up to a scalar factor. Therefore there is exactly 
one $n$-form $\omega$ normalized by the condition $\omega(
\bold E_1,\ldots,\bold E_n)=1$. It is called the volume form. For the basis 
$\bold e_1,\,\ldots,\,\bold e_n$ we have the equality
$$
\omega(\bold e_1,\ldots,\bold e_n)=\det S\cdot
\omega(\bold E_1,\ldots,\bold E_n),
$$
where the matrix $S$ is determined by the formula \mythetag{3.4}. This means 
that the value of the volume form $\omega$ can be used as a measure of 
orientation for bases in $\Bbb Q^n$.\par
   Note that if a basis $\bold e_1,\,\ldots,\,\bold e_n$ is composed
by the columns of an orthogonal matrix $O$, the transition matrix $S$ in
\mythetag{3.4} coincides with $O$. This means that constructing an orthogonal
matrix $O\in\SO(n,\Bbb Q)$ is equivalent to choosing some right oriented
ONB in $\Bbb Q^n$. One vector in this base (say last vector $\bold e_n$)
can be constructed by means of the stereographic projection as described
in section~2 above. This is the first step in our algorithm for
constructing orthogonal matrices. Then we should complement it with
other $n-1$ vectors which should be unit vectors perpendicular
to each other and perpendicular to the vector $\bold e_n$ as well. Below we 
use the Cayley transformation for this purpose.
\head
4. Cayley transformation.
\endhead
    Let $A$ be a skew-symmetric $n\times n$ square matrix. It is known
that all eigenvalues of a skew-symmetric matrix are purely imaginary
numbers (some o them can be equal to zero, but they cannot be nonzero
real numbers). Therefore $\det(1-A)$ is nonzero. Let's consider
the matrix $O=(1+A)\cdot(1-A)^{-1}$. Matrices $1+A$ and $1-A$ commute
with each other, hence $(1+A)\cdot(1-A)^{-1}=(1-A)^{-1}\cdot(1+A)$.
For this reason we write $O=(1+A)\cdot(1-A)^{-1}$ as a fraction
$$
\hskip -2em
O=\frac{1+A}{1-A}.
\mytag{4.1}
$$
The formula \mythetag{4.1} is known as the Cayley transformation (see 
\mycite{4}). If $A$ is skew-symmetric, as it is in our case, then
$O$ is orthogonal matrix with $\det O=1$. The Cayley transformation 
defines a mapping $\so(n,\Bbb Q)\to\SO(n,\Bbb Q)$ similar to the exponential
mapping $\exp\!:\so(n,\Bbb R)\to\SO(n,\Bbb R)$. But, in contrast to
the exponential mapping, it is rational, which is worth for our purposes.
The Cayley transformation is an injective mapping. Indeed, if a matrix
$O$ is obtained by means of the formula \mythetag{4.1}, one can recover 
the matrix $A$ by means of the formula
$$
\hskip -2em
A=\frac{O-1}{O+1}.
\mytag{4.2}
$$
However, one cannot apply the formula \mythetag{4.2} to an arbitrary matrix
$O\in\SO(n,\Bbb Q)$. Matrices with eigenvalues $\lambda=-1$ are not
suitable. This means that the Cayley transformation is not surjective.
Below we modify the Cayley transformation and convert it into an algorithm
able to produce each matrix $O\in\SO(n,\Bbb Q)$.\par
    We define some special skew-symmetric matrix. Taking $n-1$ rational
numbers $y^1,\,\ldots,\,y^{n-1}$, we denote by $A[y^1,\ldots,y^{n-1}]$
the matrix of the form
$$
\hskip -2em
A[y^1,\ldots,y^{n-1}]=\Vmatrix 0 &\hdots & 0 & y^1\\
\vspace{1ex}\vdots &\ddots &\vdots &\vdots\\ \vspace{1ex}
0 &\hdots &0 &y^{n-1}\\ \vspace{1ex} -y^1 &\hdots &-y^{n-1}
&0\endVmatrix.
\mytag{4.3}
$$
Substituting \mythetag{4.3} into \mythetag{4.1}, we obtain the orthogonal
matrix
$$
\hskip -2em
O[y^1,\ldots,y^{n-1}]=\frac{1+A[y^1,\ldots,y^{n-1}]}
{1-A[y^1,\ldots,y^{n-1}]}.
\mytag{4.4}
$$
By direct calculation one can find that the $n$-th column in the matrix
\mythetag{4.4} coincides with the unit vector $\bold e_n$ constructed by
means of the stereographic projection in section~2. It's components
are given by the formulas \mythetag{2.2} and \mythetag{2.3}. Let's
denote by $\bold e_1,\,\ldots,\bold e_{n-1}$ other $n-1$ columns
of the matrix $O[y^1,\ldots,y^{n-1}]$. Completed by the vector $\bold e_n$,
they form an ONB in $\Bbb Q^n$. Here are the formulas for the components
of the $k$-th vector $\bold e_k$, where $k\neq n$:
$$
\hskip -2em
\gathered
x^s=\frac{-2\,y^k\,y^s}{\left(1+\shave{\sum^{n-1}_{i=1}}
(y^i)^2\right)}\text{\ \ for \ }s\neq k,\ s\neq n,\\
\vspace{2ex}
x^k=1-\frac{2\,(y^k)^2}{\left(1+\shave{\sum^{n-1}_{i=1}}
(y^i)^2\right)},\quad
x^n=\frac{-2\,y^k}{\left(1+\shave{\sum^{n-1}_{i=1}}
(y^i)^2\right)}.
\endgathered
\mytag{4.5}
$$
We can pass to the limit $Y\to\infty$ in \mythetag{2.2} and
\mythetag{2.3}. \pagebreak However, we cannot pass to this limit in
\mythetag{4.5}. For the infinity point $Y=\infty$ we set by
definition
$$
\hskip -2em
O[\infty]=\Vmatrix 1 &\hdots & 0 & 0 & 0\\
\vspace{1ex}\vdots &\ddots &\vdots &\vdots &\vdots\\ \vspace{1ex}
0 &\hdots &1 &0 &0\\ \vspace{1ex} 0 &\hdots &0 &-1 &0\\
\vspace{1ex} 0 &\hdots &0 &0 &-1
\endVmatrix
\mytag{4.6}
$$
\mytheorem{4.1} Each unit vector $\bold e\in\Bbb Q^n$ given
by its stereographic coordinates $y^1,\,\ldots,\,y^{n-1}$ is canonically
associated with some orthogonal matrix $O\in\SO(n,\Bbb Q)$.
\endproclaim
\noindent The formulas \mythetag{2.2}, \mythetag{2.3}, \mythetag{4.5}, and
\mythetag{4.6} give an explicit proof of the theorem~\mythetheorem{4.1}. By 
construction the vector $\bold e$ is the $n$-th column in the matrix 
$O=O[y^1,\ldots,y^{n-1}]$.
\par
    Now let $O$ be some arbitrary matrix from the special orthogonal
group $\SO(n,\Bbb Q)$. We denote by $\bold e$ its $n$-th column. This
is the unit vector with rational components $x^1,\,\ldots,\,x^n$.
Suppose that $y^1,\,\ldots,\,y^{n-1}$ are its stereographic coordinates
(see formula \mythetag{2.1} in section~2). They are also rational
numbers. Therefore we can construct the orthogonal matrix $O[y^1,\ldots,
y^{n-1}]$ as described above. As a result we get two orthogonal
matrices $O$ and $O[y^1,\ldots,y^{n-1}]$ with the same $n$-th column
in them. This is possible if and only if these matrices are bound
by the relationship
$$
\hskip -2em
O=O[y^1,\ldots,y^{n-1}]\cdot\Omega,
\mytag{4.7}
$$
where $\Omega$ is an orthogonal matrix of the form \mythetag{1.2}. Thus we
have proved a theorem.
\mytheorem{4.2} Each orthogonal matrix $O\in\SO(n,\Bbb Q)$
can be represented as a product \mythetag{4.7}, where $\Omega$
is a blockwise diagonal matrix determined by some element of
the orthogonal group $\SO(n-1,\Bbb Q)$.
\endproclaim
    The theorem~\mythetheorem{4.2} is an analog of the 
theorem~\mythetheorem{1.1}, while the rational parameters $y^1,\,\ldots,
\,y^{n-1}$ are analogs of Euler angles $\varphi_1,\,\ldots,\,\varphi_{n-1}$. 
Applying this theorem recursively, for $O\in\SO(n,\Bbb Q)$ we get the 
equality
$$
\hskip -2em
O=O[y^1_1,\ldots,y^{n-1}_1]\cdot O[y^1_2,\ldots,y^{n-2}_2]
\cdot\ldots\cdot O[y^1_{n-1}].
\mytag{4.8}
$$
\mytheorem{4.3} Each orthogonal matrix $O\in\SO(n,\Bbb Q)$
can be constructed by means of the algorithm described above in the 
form of a product \mythetag{4.8}.
\endproclaim
    The theorem~\mythetheorem{4.3} is an analog of the 
theorem~\mythetheorem{1.2}. Note that the number of rational parameters 
$y^i_j$ in \mythetag{4.8} is equal to $n(n-1)/2$. This is exactly the 
same number as in the statement of the theorem~\mythetheorem{1.2}.
\head
5. Concluding remarks.
\endhead
    I am not a specialist in algebra and I am not a specialist in number
theory. I have encountered the problem of generating orthogonal matrices
with rational elements in designing computer programs for testing students 
(see \mycite{5}). Therefore it's quite possible that all of the above results 
are not new. However, I hope that being gathered in one paper and formulated 
as a computational algorithm they could be useful for practical purposes. 
In addition, I have collected some references (see \myciterange{6}{6}{--}{12}) 
related to the subject of the present paper.
\head
10. Acknowledgements.
\endhead
     I am grateful to V.~A.~Yuryev for helpful discussions. I am also
grateful to Russian Fund for Basic Research and to the Academy of Sciences 
of the Republic Bashkortostan for financial support in 2001.

\enddocument
\end